\documentclass{llncs}

\usepackage{color}
\usepackage{url}

\usepackage{booktabs}
\usepackage{ctable}
\usepackage{multirow} 
\usepackage{rotating}

\usepackage{subfigure}

\usepackage{amssymb}

\usepackage{array}

\newcolumntype{P}[1]{>{\centering\raggedright\arraybackslash}p{#1}}

\title{Web Synchronization Simulations using the ResourceSync Framework}

\author{Bernhard Haslhofer\inst{1} \and Simeon Warner\inst{2} \and Carl Lagoze\inst{3} \and Martin Klein\inst{4} \and Robert Sanderson\inst{4} \and Herbert Van de Sompel\inst{4} \and Michael L. Nelson\inst{5}}
\institute{
	University of Vienna, Faculty of Computer Science, Austria\\\email{bernhard.haslhofer@univie.ac.at}
	\and Cornell University Library, USA\\\email{simeon.warner@cornell.edu}
	\and University of Michigan, School of Information, USA\\\email{clagoze@umich.edu}
	\and Los Alamos National Laboratory, USA\\\email{mklein@lanl.gov, rsanderson@lanl.gov, herbertv@lanl.gov}
	\and Old Dominion University, USA\\\email{mln@cs.odu.edu}
	}

\begin{document}

\maketitle

\begin{abstract}
    
Maintenance of multiple, distributed up-to-date copies of collections of changing Web resources is important in many application contexts and is often achieved using ad hoc or proprietary synchronization solutions. ResourceSync is a resource synchronization framework that integrates with the Web architecture and leverages XML sitemaps. We define a model for the ResourceSync framework as a basis for understanding its properties. We then describe experiments in which simulations of a variety of synchronization scenarios illustrate the effects of model configuration on consistency, latency, and data transfer efficiency. These results provide insight into which congurations are appropriate for various application scenarios.

\end{abstract}

\section{Introduction} \label{sec:introduction}

Synchronization of resources from one Web-based system, a \emph{source}, to another, a \emph{destination}, is frequently necessary to ensure reliable access to a group of important resources, to provide backup copies for preservation purposes, or to leverage computational resources or tools available at one server but not another. The ResourceSync framework~\cite{Sompel:2012fk,Klein:2013,resyncbeta} addresses these needs within the Web architecture. It leverages the XML Sitemaps format to support baseline synchronization, incremental synchronization and audit. Example use cases include (see~\cite{wwwdev2013} for more detail):

\begin{itemize}
    \item The arXiv.org collection of articles exist on a primary server and are mirrored at other servers worldwide. The goals for synchronization of mirror sites are high consistency, moderate latency and robustness to temporary network outages. It is also desirable to make content openly available for synchonization by other services at the frequency they need without out-of-band communication to set-up the process.
    \item The data.europeana.eu~\cite{haslhofer2011data} service periodically aggregates new and changed metadata from many remote sources all over Europe. OAI-PMH is used for metadata harvesting but it would be beneficial to have a standard synchronization method for data transfer (currently manual).
    \item Structured Web data sources such as DBPedia that are synchronized with changes in their unstructured counterparts (Wikipedia). The corpora are large and undergo frequent changes so efficient incremental update mechanisms are essential.
    \item Institutional and discipline-specific repositories, and digital libraries of scholarly material based on technologies such as Fedora~\cite{lagoze2006fedora}, DSpace, etc. A standard approach that extends the current OAI-PMH metadata harvesting to include full-content sharing is required.
\end{itemize}

In all these use cases the problem can be generalized as that of \emph{synchronizing a set of changing resources from Web sources to destinations}. The synchronization problem is well-known (e.g.,~\cite{cho2003estimating}) and various solutions are available in different contexts~\cite{wwwdev2013}. However, existing approaches are either non-resource-centric or are proprietary or ad hoc solutions that are problematic when used across system boundaries.

There are aspects of synchronization that are problematic when positioned vis-\`{a}-vis the Web architecture.
Notably, synchronization of the basic component of the Web architecture~\cite{citeulike:8898564}, a \emph{resource}, cannot be formally defined because of its abstract nature. Furthermore, the concrete representations of a resource that are the outcome of dereferencing an HTTP GET request with its URI, and the content negotiation that may be part of that dereferencing process, are nondeterministic from a client-perspective; i.e., although the HTTP request may express client preferences for media type and other representation parameters, the actual representation returned has no identifier, is server-determined, and may be spatially, temporally~\cite{van2009memento}, personally, and user-agent dependent. Therefore, the notion of synchronization and a formal model for it can only be expressed within a constraint on the Web architecture with limited, deterministic, and URI-identified representations for resources.


This paper defines synchronization within this constrained but important context and makes the following contributions:

\begin{enumerate}

    \item We formally define a synchronization model and the constraints on the Web architecture that support it. In terms of this model, we describe the modular set of ResourceSync synchronization components for baseline and incremental synchronization that can be flexibly combined to meet a variety of requirements.

    \item We explain the results of our experiments that illustrate the effects of possible model configurations on consistency, latency, and data transfer efficiency, providing intuition on which configurations are appropriate for various scenarios.
    
\end{enumerate}

A simulator, tools and libraries used for this work are open source and available at \url{https://github.com/resync}.




\section{Resource Synchronization}\label{sec:res_sync}

In this section, we first review some of the basic concepts of the Web Architecture, and then explain the constraints we are imposing to define and implement resource synchronization. We then present our general model and framework for resource synchronization, and explain how this approach is implemented in ResourceSync.

\subsection{Resource-centric Synchronization}

The Web architecture~\cite{citeulike:8898564} distinguishes between \emph{identifiers}, \emph{resources}, and resource \emph{representations}: resources are identified by URIs and dereferencing a URI via HTTP yields a representation of the current state of the resource. URIs are opaque, meaning that resource state may evolve over time without requiring a URI owner to republish a new URI for each change in a resource state. Thus, the Web architecture promotes independence between an identifier (URI) and the state of the identified resource, which is defined by its representations.

A resource can have multiple representations and content negotiation\footnote{\url{http://www.ietf.org/rfc/rfc2616.txt}} allows clients to request from a negotiable resource the nature of the representation it wants by expressing preferences in special-purpose accept headers. Example representations of a single resource might include: a document in HTML and English, a document in PDF and Chinese, and a raw data representation of the same resource in JSON. However, the mapping from resource to representations is, from a client-perspective, non-deterministic because the output of the dereference function depends on external state. A server could, for instance, return different representations depending on the clients location, user agent, browsing history as recorded by cookies, or even randomly.

\subsection{Constraints and Basic Definitions}\label{sec:constraints}

We apply the following constraints to the Web Architecture to make it possible to formally define and implement resource synchronization:

\begin{enumerate}

    \item We define the notion of a \emph{synchronization context} in which user agents identify their synchronization-specific HTTP requests as taking place within that context and servers recognize the requirements of that context. Here we follow a design paradigm that is known from Web crawling and which allows servers to construct resource representations dedicated to that process~\cite{Brandman:2000:CWS:362883.362894}. We denote Web servers that implement these constraints as \emph{sources} and application environments that implement ResourceSync clients as \emph{destinations}.

    \item Within the synchronization context, a resource to be synchronized must map to a single representation. This does not limit content negotiation, but requires it to be implemented as Transparent~\cite{holtman1998transparent} or 303 style~\cite{Heath:2011kx} content negotiation, whereby each representation has its own URI. This means that synchronization is supported only for information resources \emph{whose essential characteristics can be conveyed in a message}~\cite{citeulike:8898564}.
    
    \item Within the synchronization context, we require representations constructed by a source to be comparable, meaning that an \emph{equals} function applied on the payload of two independently requested resource representations returns \emph{true} if the entity body of the HTTP response (e.g., the served PDF document) is byte-equivalent and false otherwise.

\end{enumerate}

\begin{figure}
    \centering
    \includegraphics[width=0.6\textwidth]{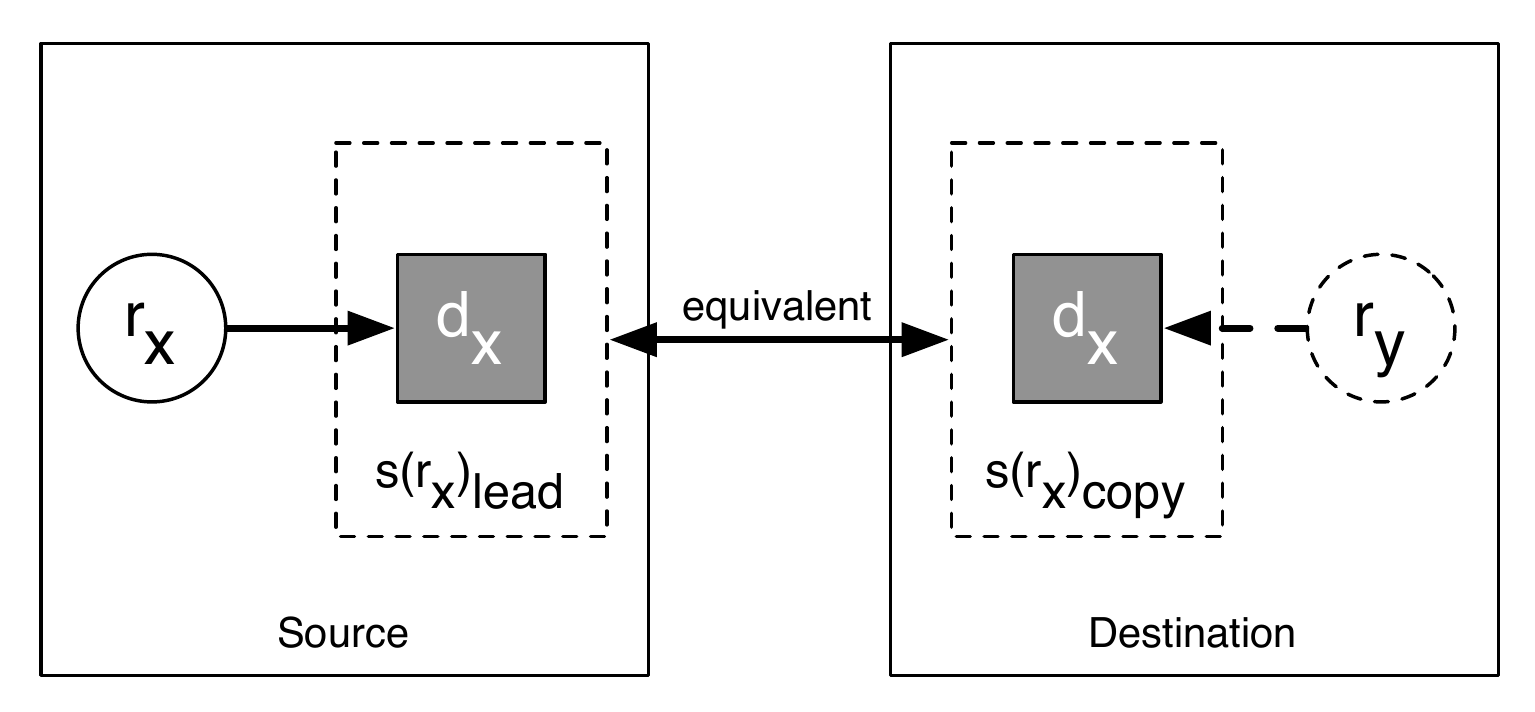}
    \caption{In-sync-ness of a single resource pair.}
    \label{fig_in_sync_ness}
\end{figure}

Given these constraints and definitions, we can now refine the resource synchronization problem as follows: if we let $r=\{u,d\}$ be a Web resource, with $u$ being its URI identifier and $d$ being an associated, distinguishable representation, we can define the state of that resource as $s(r)=d$. To synchronize a single resource from a source, a destination must create a \emph{copy state} $s_{copy}$ from a resource's \emph{lead state} $s_{lead}$ at the source. These resource states are said to be \emph{in sync} if $s_{copy} \equiv s_{lead}$. Figure~\ref{fig_in_sync_ness} illustrates this notion for a destination holding a single copy state which is in sync with its lead state. Note that the copy state at the destination may be stored in a closed environment (e.g., file system) or could be associated with another Web resource $r_y$.

%

A destination is \emph{consistent} with a source if all copy resource states at the destination are in sync with their corresponding lead states at the source. A destination is \emph{inconsistent} with a source if at least one copy state is not in sync with the corresponding lead state. Inconsistencies can occur either when a resource state changes at the source but not at the destination, and vice versa.

A \emph{change} affects a resource so that $s_{t-1}(r) \neq s_t(r)$. It can be described as $c=\{u_r,type,t\}$, with $u_r$ being the identifier of the affected resource $r$, $type \in \{create, update, delete\}$ expressing change semantics, and $t$ being --- as in HTTP Last-Modified --- the date and time at which the source claims a resource representation was changed.

We define \emph{synchronization} as an operation that aims to keep a destination synchronized with changes in a source by updating copy states of resources to be in sync with their lead states. Technically, this can be achieved through dereferencing the corresponding source resource for each given copy state $s_{copy}$ and replacing it by its lead state $s_{lead}$. Destinations repeatedly synchronize with a source in order to maintain consistency as changes occur over time.

\subsection{Framework Components}

In order to allow a destination to initially synchronize with a source, the synchronization process at the destination must be able to retrieve a list of source resources for which synchronization is intended --- we denote this process as \emph{baseline synchronization}. Subsequently it can perform \emph{incremental synchronization} to keep its copy states in sync with the corresponding resources' lead-states. These two functionalities are implemented via \emph{resource lists} and \emph{change lists}, described in the sections below.

\subsubsection{Resource List}

Baseline synchronization requires that a source exposes a resource list. In its most basic form, it comprises a list of URIs, each of which, when being dereferenced, returns exactly one possible representation. If a source also exposes information about the time of the latest lead state change, i.e., a last-modified timestamp for each resource, destinations can restrict synchronization to the set of resources that carry a timestamp greater than the time of the destination's previous synchronization cycle. The source can also compute a hash digest for this resource and expose it as part of the inventory to save the destination from comparing possibly unchanged resource states. Formally, a resource list can be defined as a set of entries $RL =\{e_1, e_2, \ldots, e_n\}$, where each entry $e_i = \{u_r, lm_d, hash_d, size_d\}$ contains the identifier $u_r$ of a resource $r$ and metadata that is computed over its representation, such as last modified date ($lm_d$), hash digest ($hash_d$), and byte-size ($size_d$).

A resource list is a Web resource on its own and can be regarded as snapshot of a source's resource collection at a certain point in time. The size of a resource list is proportional to the number of resource states exposed by a source and affects its transfer time. Large resource lists thus increase latency and, as a consequence, the time it takes for a destination to achieve consistency with a source. We describe these effects more fully in the experiments section.

Destinations implement baseline synchronization as follows: synchronization is initiated by a destination, which fetches a resource list via HTTP. It then iterates through all entries in this list and checks for each entry if a corresponding local copy state exists. If not, then that lead state is fetched from the source via HTTP. If a copy state is available, but no corresponding lead state can be found, then the destination should delete the local copy.








\subsubsection{Change List}

Incremental synchronization can reduce latency caused by the transfer of possibly large resource lists, and is therefore an optimization and optional component in the proposed synchronization framework. Instead of retrieving the list of available resources, destinations can retrieve atomic resource state change information bundled in change lists.

A change list represents an ordered list of changes $CL=\{c_1, c_2, \ldots, c_n\}$, where each change entry $c_i$ carries at least information about the affected resource and the time and type of a change (see Section~\ref{sec:constraints}). In combination with at least one previously retrieved inventory and zero or more change lists, it allows destinations to reproduce source resource state in an atomic manner.



Incremental synchronization can be executed by a destination after at least one baseline synchronization iteration: after fetching a change list it chronologically iterates its entries and either creates, updates, or deletes local copy states.


Synchronization performance can be increased by additional components that optimize specific aspects of the synchronization process. \emph{Resource dumps}, for instance, could be exposed by a source and save destinations from executing a possibly high number of HTTP requests against listed resources. A dump is essentially an archive (e.g., ZIP file) containing a resource list and packaged representations, which correspond to the state of the resources listed in the resource list at the time of the output of the dump. Destinations can download dumps, unpack and apply them just as they would apply resource lists, with the major difference that copy states are created by extracting files from an archive instead of dereferencing the corresponding resources' URIs. In a similar manner, resource states could also be packaged with change lists in \emph{change dumps}.



The ResourceSync specification~\cite{resyncbeta} is a concrete implementation the described synchronization approach. It leverages widely adopted XML Sitemaps~\cite{schonfeld2009sitemaps} for expressing resource lists, change lists, resource dumps, change dumps and also proposes notification components that can be combined by implementers depending on a source's capabilities. ResourceSync clients can discover capabilities via a central lookup file that is accessible from a source's Web root.


\section{Experiments}\label{sec:experiments}

We report here on a series of simulations we conducted to illustrate the effects of possible model configurations by varying source- and destination-specific characteristics. We use the following metrics:

\begin{enumerate}
 
    \item \emph{Average Consistency}:
    We define consistency at a given time $t$ as the fraction of resource copy states that are in sync with their lead states. The average consistency of a destination w.r.t  a source in a given time interval is then the mean consistency over the entire interval.

    \item \emph{Average Latency}: Latency is defined as the time it takes to achieve in sync state at the destination after a resource representation in the source changed. We can compute latency for each $s(r)_{lead}$ and $s(r)_{copy}$ pair and compute average latency for the same time interval.

    \item \emph{Data Transfer Efficiency}:
    Retrieving meta-information about resources to be synchronized generates additional data transfer overhead and, as a consequence, latency. We measure this aspect by computing data transfer efficiency, which is defined by the fraction of ``required bytes'' in the ``total bytes'' transferred in a single simulation iteration.

\end{enumerate}



All simulations were conducted using the ResourceSync Simulator\footnote{\url{https://github.com/resync/simulator}}, which simulates a source and generates synthetic resources, each having exactly one non-negotiable representation of random size. We deployed our simulation environment on pairs of Amazon EC2 instances, one instance simulating a source, the other simulating a destination.
We simulated two different types of sources: one frequently changing source with a change interval of 0.1 seconds and one less frequently changing source with a 10 seconds change interval. The simulations were configured with varying numbers of resources (100, 1k, 10k, 25k, 50k) and varying destination-triggered synchronization intervals (10 sec, 100 sec).
In all cases the maximum representation size was set to 1k bytes.


\subsection{Results}

\newlength{\fw}
\setlength{\fw}{0.8\textwidth}
\begin{figure}
  \centering
  \includegraphics[width=\fw]{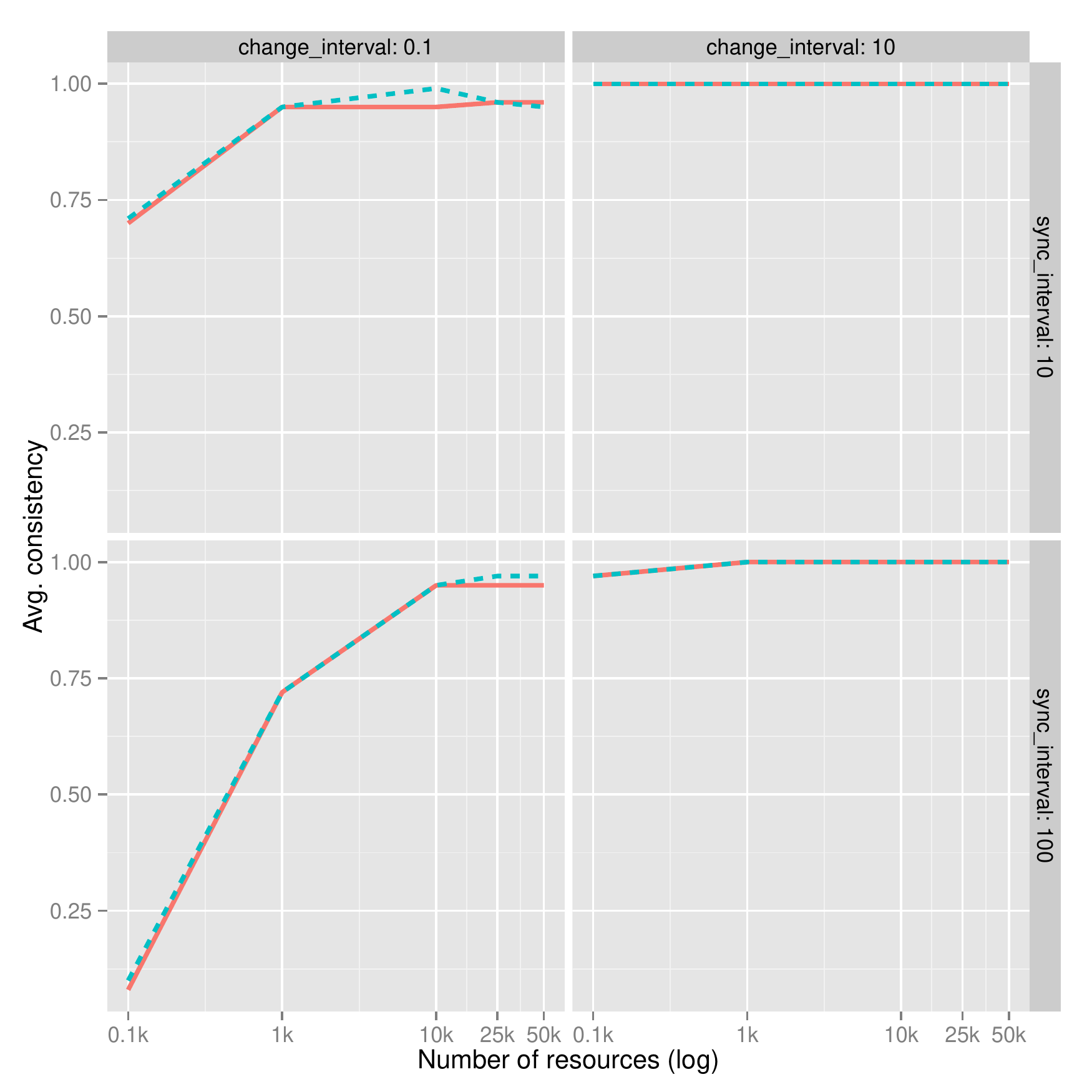}
  \caption{\label{fig_avg_consistency}Average consistency with varying numbers of resources in baseline (red/solid) and incremental (blue/dashed) synchronization. Plots with 0.1 and 10~s change interval, and 10 and 100~s synchronization interval.}
\end{figure}

Figure~\ref{fig_avg_consistency} shows that synchronizing frequently (10 sec) with a small (100 resources) and rapidly changing (0.1 sec change interval) source leads to higher average consistency than less frequent synchronization intervals (100 sec). This effect depends on the fraction of changes that occurred at the source within a single synchronization interval: smaller sources with high change frequency have a higher fractional change than larger sources, which explains why this effect disappears in larger sources. Incremental synchronization does not increase average consistency, because the additional effort caused by repeatedly downloading resource lists for baseline synchronization is marginal compared to the time it takes to synchronize the resource representations from source to destination.


\begin{figure}
  \centering
  \includegraphics[width=\fw]{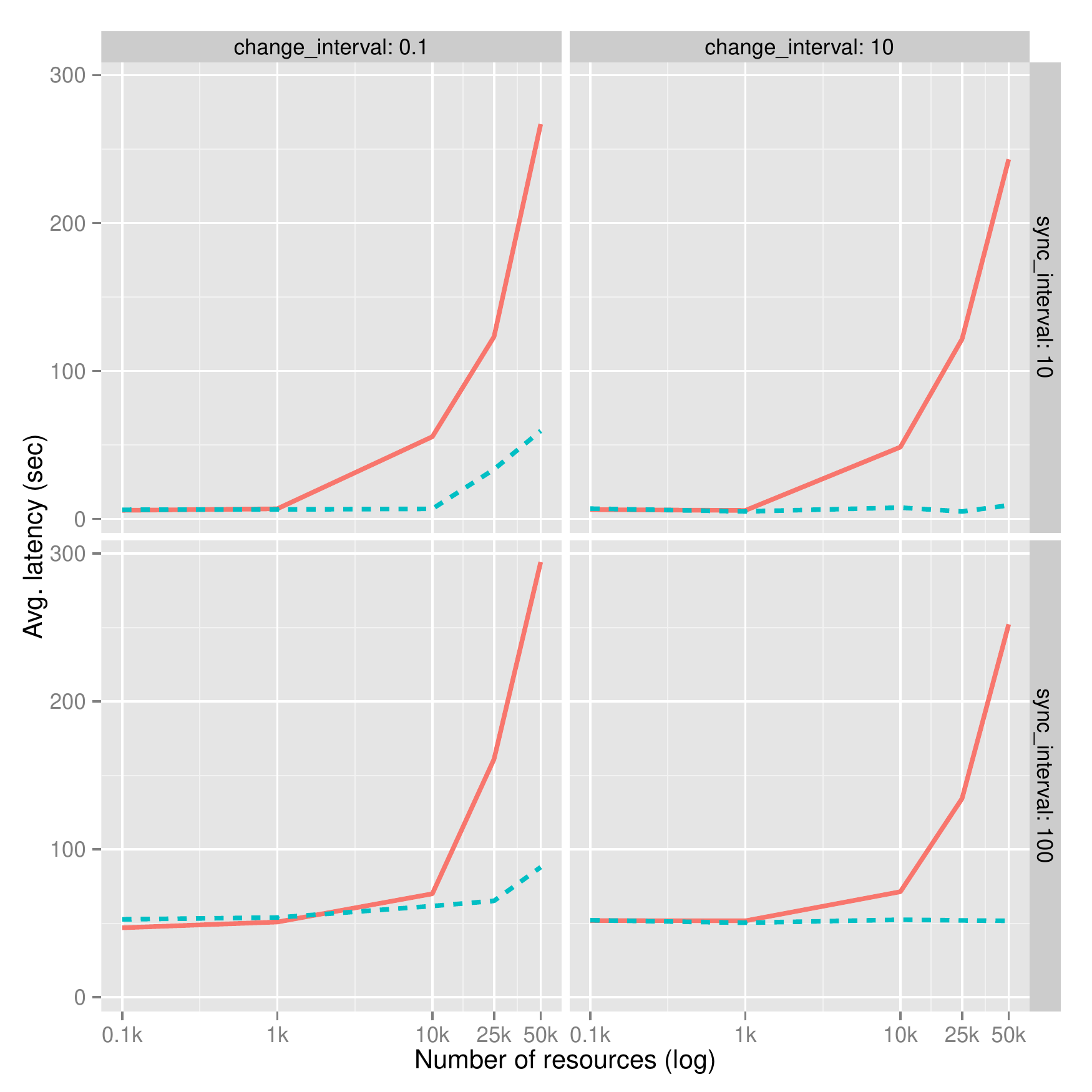}
  \caption{\label{fig_avg_latency}Average latency with varying numbers of resources in baseline (red/solid) and incremental (blue/dashed) synchronization. Plots with 0.1 and 10~s change interval, and 10 and 100~s synchronization interval.}
\end{figure}

Figure~\ref{fig_avg_latency} shows that the average time it takes for a copy state at a destination to get in sync with its lead state increases rapidly with a growing number of resources and leads to high average latency in baseline synchronization mode. This is because the transfer of the resource list itself becomes a significant delaying factor. In our simulation environment it is clear that a source with more than 1-10k resources should expose change lists if they need to support low-latency synchronization on the destination-side.

\begin{figure}
  \centering
  \includegraphics[width=\fw]{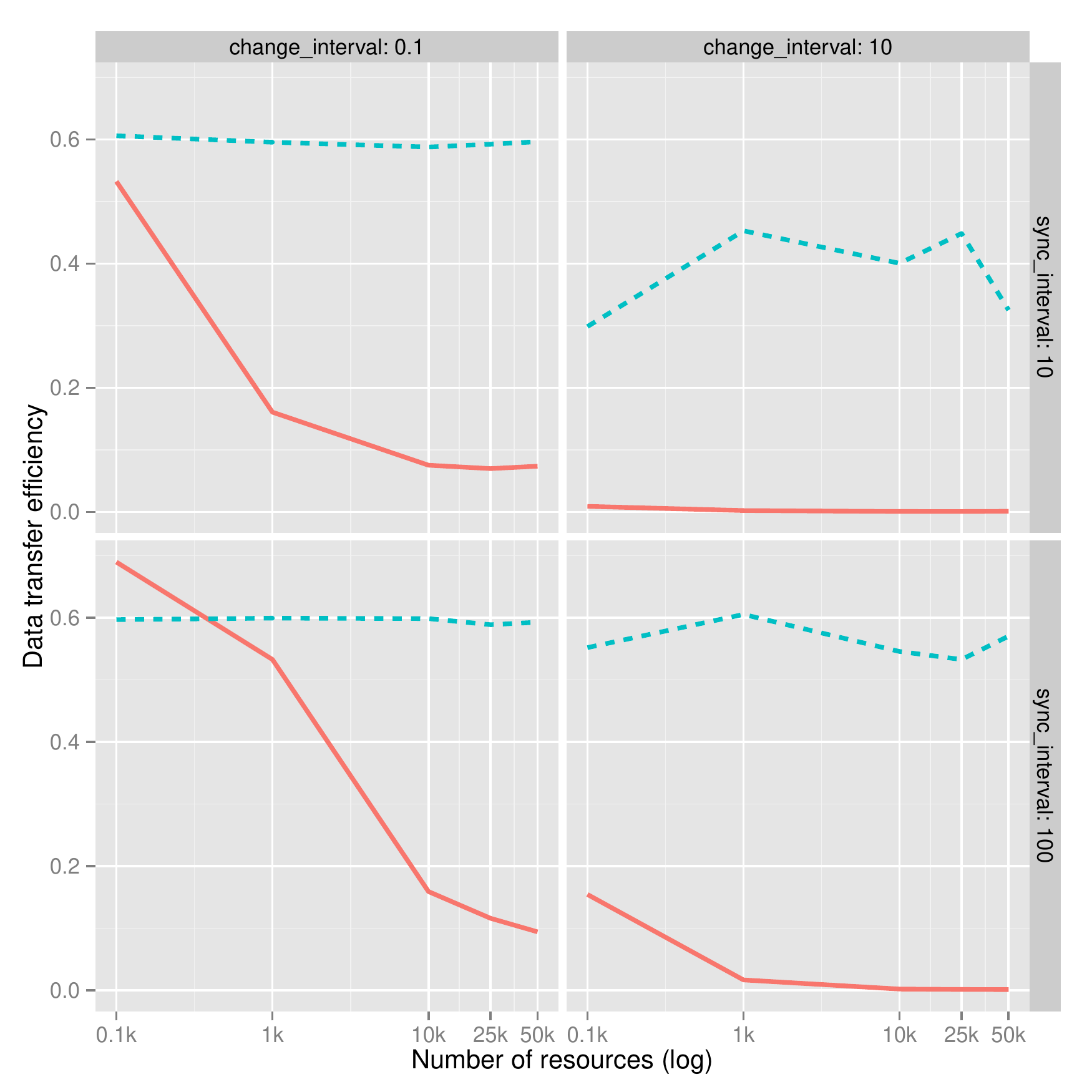}
  \caption{\label{fig_data_transfer_efficiency}Data transfer efficiency with varying numbers of resources in baseline (red/solid) and incremental (blue/dashed) synchronization. Plots with 0.1 and 10~s change interval, and 10 and 100~s synchronization interval.}
\end{figure}

Baseline synchronization becomes less efficient with increasing total number of resources because the data transfer overhead is proportional to the number of resources. The efficiency when using change lists is approximately constant because the overhead is proportional to the number of changed resources that must be transferred. If a source supports only baseline synchronization and just a small fraction of the total number of resources change between synchronization events, then destinations polling resource lists will see significant overhead when compared against transferring change lists only.
Figure~\ref{fig_data_transfer_efficiency} shows this difference in the upper and right graphs (the line for change list synchronization in the upper-right graph has statistical noise because there were a small number of changes in the simulation period).
The situation is reversed for the left side of the lower-left graph where the fraction of changed resources is high and the data overhead caused by the resource list is low. In this regime exposing change lists does not increase data transfer efficiency.

\subsection{Observations}

When destinations synchronize with a source, the goal is typically to achieve high consistency and low latency. We showed that supporting incremental synchronization in smaller sources does not lead to higher consistency and lower latency because the resource lists exposed by a source are reasonably small and can quickly and efficiently be downloaded and processed by destinations. Thus the level of achievable average consistency mainly depends on the destination's synchronization interval: if the source changes frequently, a destination needs short synchronization intervals, if it changes infrequently, longer synchronization intervals are adequate for high consistency. Since there is a linear relationship between the number of source resources and the size of a source's resource list and therefore also its download time, average latency rises with an increasing number of source resources. This means that incremental synchronization, enabled by change lists, can reduce average latency especially when sources are synchronizing with larger destinations. A similar effect is observable w.r.t data transfer efficiency: the larger resource lists in proportion to the number of changes, the less efficient it is to transfer them entirely to a destination. Hence, for small sources it is sufficient to expose resource lists only, while larger, frequently changing sources should also expose change lists to enable high-consistency and low-latency synchronization between sources and destinations.

Our observations can be summarized as follows and serve as configuration guidelines for adopters of ResourceSync or other instances of our Web-based synchronization model:

\begin{itemize}
    \item High consistency requires rapid change notification, which can be achieved by short destination-side synchronization intervals and the use of change lists. (This regime is also well suited to push-based notification which is supported by ResourceSync but not discussed in this paper.)
    
    \item If a source exposes a small number of resources it is sufficient to support resource lists only. Support for change lists does not have a positive effect on consistency, latency, or data transfer efficiency.
    
    \item Any medium- to large-size source should implement change lists to reduce average latency and optimize data transfer efficiency.
    
\end{itemize}

\section{Related Work}\label{sec:related_work}



The synchronization problem has been discussed in the context of clock synchronization~\cite{lamport1978time}, concurrency in distributed databases (e.g.~\cite{bernstein1981concurrency}), large-scale clould computing (e.g.,~\cite{decandia2007dynamo}). ResourceSync is certainly not designed for that purpose but focuses on global synchronization of resources across system boundaries.

The problem of changing resources has been discussed by Cho and Molina~\cite{cho2003estimating} for general Web documents and by Umbrich et al.~\cite{umbrich2010towards} for Linked Data resources. Linner~\cite{linner2012instant} gives an overview of existing techniques and proposes an instant state synchronization approach for Web hypertext applications. His work is closely related to ours, with the main difference being that ResourceSync does not target real-time synchronization of application events. 

A variety of tools has been proposed to synchronize resources in distributed systems: \emph{rsync}~\cite{tridgell1996rsync} allows the synchronization of files between file systems. This closely reflects our synchronization approach, with the main differences being that rsync (i) is not Web-based and (ii) requires that sources run a rsync server that supports differential update computation. \emph{Collection Synchronization for WebDAV}\footnote{\url{http://tools.ietf.org/html/rfc6578}} supports the synchronization of contents in WebDAV collections and provides baseline synchronization functionality. However, it introduces WebDAV-specific HTTP verbs (e.g., PROPFIND, REPORT).
The \emph{Atom Syndication Format}\footnote{\url{http://www.ietf.org/rfc/rfc4287}} allows sources to expose the entire content or changes in Web sites (e.g., blogs) as Web feeds. However, it contains mandatory metadata elements (e.g., summary, author) that would increase the data transfer overhead for non-document Web resource synchronization. \emph{OAI-PMH}\footnote{\url{http://www.openarchives.org/pmh/}} was designed for metadata harvesting. ResourceSync should not necessarily replace OAI-PMH but provide a standardized, and toolset-supported mechanism for synchronizing also the resources.


\section{Conclusions}\label{sec:conclusions}

In this paper, we addressed the problem of synchronizing a collection of changing resources from Web sources to destinations. We outlined a conceptual model for resource-centric synchronization and defined three constraints to the existing Web architecture that are necessary for defining and implementing synchronization between a source and a destination.
We reported on a number of simulations that illustrate the effects of model configuration parameters on average consistency, average latency, and data transfer efficiency. Our results show the effectiveness of implementing change lists for larger, frequently changing sources, whereas implementing inventories is sufficient for smaller sources, independent of their change frequency.




\bibliography{references}
\bibliographystyle{splncs03}

\end{document}